\documentclass{iopart}
\usepackage{color}
\newtheorem{theorem}{Theorem}
\newtheorem{lemma}{Lemma}

\usepackage{graphicx}

\begin{document}

\newcommand{\be}{\begin{equation}}
\newcommand{\ee}{\end{equation}}
\newcommand{\beqn}{\begin{eqnarray}}
\newcommand{\eeqn}{\end{eqnarray}}
\newcommand{\su}{\mbox{\rm su}}

\newcommand{\he}{H_{\rm e}}

\title{Complete controllability of finite quantum systems with
two-fold energy level degeneracy}

\author{Zhedong Zhang and H.\,C.\,Fu\footnote{E-mail: hcfu@szu.edu.cn}}

\address{School of Physical Sciences and Technology, Shenzhen
University, Shenzhen 518060, P.\,R.\,China}

\begin{abstract}
Complete controllability of finite dimensional quantum systems
with energy level degeneracy is investigated using two different
approaches. One approach is to apply a weak constant field to
eliminate the degeneracy and then control it using techniques
developed for non-degenerate quantum systems. Conditions
for the elimination of degeneracy are found and the issue
of influence of relaxation time of constant external field to
the target state are addressed through the fidelity.
Another approach is to control the degenerate system by a single
control field directly. It is found that the system with
two-fold degenerate excited states and non-degenerate ground
state are completely controllable except for the two-level
system. Conditions of complete controllability are found for
both systems with different energy gaps and with equal energy
gaps.
\end{abstract}

\maketitle

\section{Introduction}

Quantum control is essentially understood as a coherent or incoherent manipulation
of a quantum system, which attempts a time evolution from an arbitrary initial
state to an arbitrarily given target state \cite{1,2,3,4}. It was first
proposed by Huang {\it et.\,al.} in 1983 \cite{huang} and was then
further developed for application in
the control of chemical reactions. In recent years, the
controllability of quantum system has been well
defined \cite{controllability,chem,solo} and investigated using algebraic methods
\cite{fu1,fu2}, graph methods \cite{graph1,graph2} and transfer graph
methods \cite{graph3}. Cabrera {\it et.\,al.} investigated a sufficient
condition for the {\it state control} of symmetric degenerate quantum systems
and found that the underlying Lie algebra could be the symplectic algebra rather than
su($N$) \cite{cab}.
However, it seems that the
control of quantum systems with energy level degeneracy has not been investigated
systematically in the literature and it is obviously significant
to investigate quantum control of systems with energy level
degeneracy.

In this paper, we shall present two different approaches to the
controllability of finite dimensional quantum system
with two-fold degenerate excited states and non-degenerate
ground state. One approach is to first eliminate
the degeneracy using a constant excitation field and then control it
using the techniques developed for the control of non-degenerate system.
In this approach one has to choose the proper external excitation
field such that the energy structure of the degeneracy-eliminated
system meets the controllability
criteria for the complete controllability of non-degenerate system
\cite{fu1,fu2}.
In quantum control we generally require that the control fields attenuate
to zero when the target state is achieved. However, the excitation
field is a constant field and we cannot switch it off instantly when
the target state is achieved, rather the excitation field tends to
zero in a period of time (relaxation time) and this will definitely
cause a shift of target state. We shall examine how much the final
state shifted from the target state by evaluating their fidelity.
We find that the shift of final state from the target state is minor
if the decay time is short enough.

Another approach is to use a single external control field to control the
degenerate system directly without eliminating energy degeneracy first.
Complete controllability is systematically investigated using Lie algebra method.
Lie algebra method is an important method for the investigation
of both classical \cite{liec} control theory and quantum control theory
\cite{fu1,fu2,fu3}.
It is found that the degenerate systems can be completely controlled
through a single control field if some conditions are
fulfilled except the simplest two energy level system.
Controllability conditions are algebraically found.

This paper is organized as follows. In Sec.\,II we formulate
two control schemes and fix some notations. In Sec.\,III we
investigate the control of degenerate system by eliminating the degeneracy
using constant excitation field. In Sec.\,IV we turn to the
direct control of degenerate system using single control field. We conclude in
Sec.\,V.

\section{Control of the system with energy degeneracy}

Consider an $N$-level quantum system described by the following Hamiltonian
\begin{equation}
{H_0}={\sum_{n=1}^N}{\sum_{k=1}^{\beta_n}} E_n|n,k\rangle\langle n,k|,
\label{system}
\end{equation}
where $E_n$ is the eigen energy of the $n$-th level and ${|n,k\rangle}$
are corresponding eigenstates. Here we only consider the case where
the ground state is non-degenerate and all excited states are two-fold
degenerate, namely
\begin{equation}
\beta_n=\left\{
\begin {array}{ll}
    1, & \mbox{when }n=1;\\
    2, & \mbox{when }n\geq 2.
\end {array}
\right.
\end{equation}
Our aim is to steer the system (\ref{system}) to an arbitrary given target
state by interacting with classical fields. As the controllability of
finite systems without degeneracy has been extensively studied \cite{controllability,
fu1,fu2}, a natural idea is to apply a constant excitation field to eliminate the
energy degeneracy and then control it using the techniques developed
for the non-degenerate system.
Suppose that the interaction Hamiltonian $\he$ between the excitation field
and the system takes the following form
\begin{eqnarray}
  \he=&\sum_{n=1}^{N-1} \sum_{k=1}^{\beta_n}
        \sum_{p=1}^2 g_{nk,n+1p}\left(|n,k\rangle\langle n+1,p|+|n+1,p\rangle\langle n,k|\right)   \nonumber \\
      & +\sum_{n=2}^N g_{n1,n2} \left(|n,1\rangle\langle n,2|+|n,2\rangle\langle n,1|\right),
\end{eqnarray}
where $g_{mn,pq}$ are real constants.
Then apply the control fields to control the excited system. In this paper,
we are particularly interested in the case of a single control field. In this case
the interaction Hamiltonian takes the following dipole form
\begin{eqnarray}
  H_I=& \sum_{n=1}^{N-1} \sum_{k=1}^{\beta_n}
  \sum_{p=1}^2 d_{nk,n+1p}(|n,k\rangle\langle n+1,p|+|n+1,p\rangle\langle n,k|).
  \label{4}
\end{eqnarray}
Then the total Hamiltonian of the control system is
\be
H=H_0+\he + f(t)H_{I},
\ee
where $f(t)$ is the classical control field.
For this scheme we need to address two issues:
\begin{description}
\item{(1)} After the degeneracy is eliminated, whether the
degeneracy-eliminated system meets the controllability criteria given in the
\cite{fu1,fu2};
\item{(2)} In quantum control, we generally require that the control
fields approach to 0 when the control time $T$ is reached and the target
state is achieved. But for the constant excitation field removing the
degeneracy, we have to turn it off when the target state is archived and
this relaxation needs time which may cause the target state a shift. So
we need to answer how the relaxation time affect the target state.
\end{description}
We will address both issues in Sec.\,\ref{section3}.

Another approach is to control the system (\ref{system}) using the control
field (\ref{4}) directly, just like the control of non-degenerate system.
In this case the Hamiltonian of the total control system is
\begin{equation}
H=H_0+ f(t)H_I.
\end{equation}
We will see in Sec.\,\ref{section-control-2} that the system can be completely controlled
if some conditions about the coupling constants in $H_I$ are satisfied
except the simplest two energy level system.

Let us fix some notations we will use hereafter.
For convenience, let us denote $e_{ij,kl} = |i,j\rangle\langle k,l|$
and define the following skew-Hermitian operators
\begin{eqnarray}
x_{nk,ml} = i\left(|n,k\rangle\langle m,l|+|m,l\rangle\langle n,k|\right), \quad n<m,\nonumber \\
y_{nk,ml} = |n,k\rangle\langle m,l|-|m,l\rangle\langle n,k|, \quad n<m, \nonumber \\
h_{nk,ml} = i\left(|n,k\rangle\langle n,k|-|m,l\rangle\langle m,l|\right),\quad n<m,
\label{su2n-1}
\end{eqnarray}
on the Hilbert space of the considered system with dimension $2(N-1)+1=2N-1$.
Those operators generate the Lie algebra $\su(2N-1)$.
To prove the complete controllability, we need to prove that the Lie algebra $L_0$ generated
by $iH_0$ and $iH_I$ is  $\su(2N-1)$, or in other words, generate all
operators in (\ref{su2n-1}). In fact, we only need to
prove that $iH_0$ and $iH_I$ generate the following operators
\be
x_{nk,n+1\,l}, \quad
y_{nk,n+1\,l}, \quad
h_{nk,n+1\,l}, \quad
1\leq n \leq N-1;\ k,l=1,2,\label{8}
\ee
as from those operators we further have
\begin{eqnarray}
\left[x_{nk,n+1\,l}, y_{n+1\,l, n+2\,p}\right] = x_{nk,n+2\,p},\nonumber\\
\left[y_{nk,n+1\,l}, y_{n+1\,l, n+2\,p}\right] = y_{nk,n+2\,p}, \nonumber \\
-2^{-1}\left[x_{nk,n+2\,p}, y_{nk, n+2\,p}\right] = h_{nk,n+2\,p},
\end{eqnarray}
and then all elements in (\ref{su2n-1}) recurrently. Therefore the system
is completely controllable if elements (\ref{8}) can be generated by
$iH_0$ and $iH_I$.

\section{Control by elimination of degeneracy}\label{section3}

Let us first briefly discuss
conditions of complete controllability of degenerate systems
by eliminating degeneracy through a constant classical field.
Suppose that the excitation field is week enough that we can use the perturbation
theory to evaluate the new energy structure of the system.
It is easy to find the first order approximation of the eigen energy
\begin{equation}
E_{nk}={E_n}+{E_{nk}^{(1)}},
\end{equation}
where $E_{n1}^{(1)}=-|g_{n1,n2}|$ and $E_{n2}^{(1)}=|g_{n1,n2}|$.
As the excitation field is week, we can require the degeneracy-removed
system does not have energy level crossing
\be
    E_{n+1,1}>E_{n,2}.\label{hello1}
\ee
To meet the controllability criteria for non-degenerate system given in
\cite{fu1,fu2}, we require
that, for example, the first energy gap
$(E_2-\Gamma_2)-E_1$ is different from any others, namely
\begin{eqnarray}
& ({E_2}-\Gamma_2)-E_1  \neq (E_n +\Gamma_n)-(E_n -\Gamma_n)=2 \Gamma_n, \nonumber\\
& ({E_2}-\Gamma_2)-E_1  \neq
(E_n -\Gamma_n)-(E_{n-1}+\Gamma_{n-1}), \nonumber \\
&  \qquad (n=2,3,...,2N-1), \label{hello2}
\end{eqnarray}
by proper choice of the coupling constants $g_{n1,n2}$. The system
is thus completely controllable if conditions (\ref{hello1}) and
(\ref{hello2}) are fulfilled.

We then address the issue  of the influence of relaxation time to
the target state.
Assume that the system has been driven to the normalized target state at time $T$
\be
 |\psi{(T)}\rangle =
\sum_{n=1}^N \sum_{p=1}^{\beta_n} C_{np} (T)|n,p\rangle.
\ee
We then switch off the constant
excitation field and, without losing generality,
we assume that this process is governed by the Hamiltonian
\be
H=H_0+e^{-(t-T)/{\tau}}H_{e}, \quad t\geq T,
\ee
where $\tau$ is the relaxation time.
From standard time-dependent perturbation theory, we can easily obtain
the the state of the first order at the {\it time of half decay} $T_e=\tau \ln2+T$
\be
|\psi_I(T_e)\rangle=\sum_{m=1}^{N} \sum_{k=1}^{\beta_m}
\Big(C_{mk}(T)+C_{mk}^{(1)}(T_e)\Big)|m,k\rangle,
\ee
where
\beqn
\fl\qquad
C_{mk}^{(1)}(T_e)
=
\sum_{p=1}^{\beta_{m-1}} E_{m,m-1}(\tau) \left(1-\frac{1}{2}e^{i\omega_{m,m-1}\tau\ln{2}}\right)
C_{m-1,k}(T)g_{mk,m-1p} \nonumber\\
\fl\qquad\qquad\qquad + \sum_{p=1}^{\beta_n}\frac{\tau}{2i\hbar}C_{mp}(T) g_{mk,mp}   \nonumber \\
\fl\qquad\qquad\qquad + \sum_{p=1}^{\beta_{m+1}}E_{m,m+1}(\tau)\Big(1-\frac{1}{2}
e^{i\omega_{m,m+1}\tau\ln{2}}\Big)C_{m+1,k}(T) g_{mk,m+1p}.
\eeqn
and $\omega_{mn}=(E_m-E_n)/{\hbar}$, $g_{mk,np} =\langle m,k|H_{e}|n,p\rangle$,
$E_{m,m\pm 1}(\tau)=1/(E_m-E_{m\pm 1}+i\hbar/\tau)$.

To compare the difference between target state and $|\psi_I(T_e)\rangle$, we
examine the fidelity $F$ between two states
\be
F\equiv \frac{|\langle\psi(T)|\psi_I(T_e)\rangle|^2}{
\langle\psi(T)|\psi(T)\rangle\langle\psi_I(T_e)|\psi_I(T_e)\rangle}
=\frac{|\langle\psi(T)|\psi_I(T_e)\rangle|^2}{
\langle\psi_I(T_e)|\psi_I(T_e)\rangle}.\label{fidelity1}
\ee
It is obvious that in the ideal case $\tau\to 0$, $C_{mk}^{(1)}(T_e)\to 0$
and thus $F\to 1$, as we expected. As the excitation field  is
week, one generally has $|H_{mk,np}/(E_m-E_n)|\ll 1$. So
if the relaxation time $\tau$ is short enough (This means generally it is
much less than the characteristic time of the atom, $\sim 10^{-11}\mathrm{s}$),
$C_{mk}^{(1)}$ is small and the
fidelity between the target state and final state is close to 1.

\begin{figure}
\includegraphics[height=7cm]{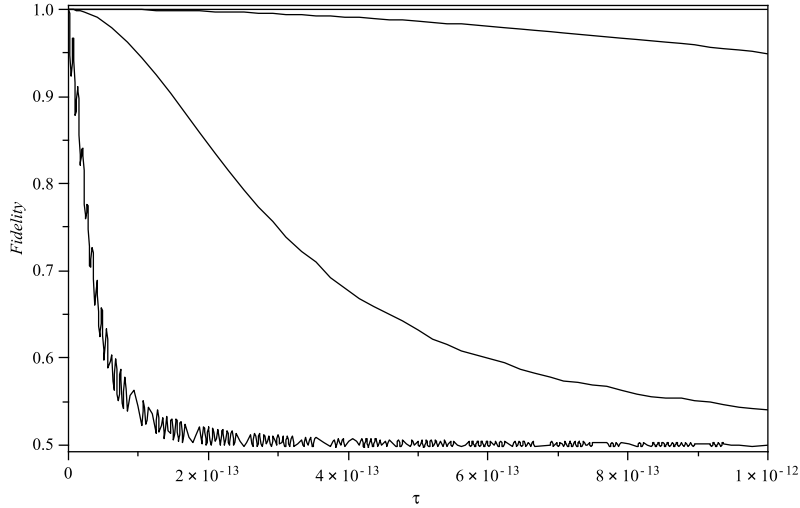}
\caption{Fidelity against relaxation time $\tau$. The energy gap $E_2-E_1$
is chosen as 1eV and $C_{11}, C_{21}, C_{22}$ are chosen as $1/\sqrt{2},1/2,1/2$
respectively. Corresponding to four curves from top to bottom, we take
matrix elements
$g_{11,21}= g_{11,22}$ which are chosen as $10^{-23}, 10^{-22}, 10^{-21}$ and $10^{-20}$,
respectively. }
\label{fig1}
\end{figure}

In Fig.\ref{fig1}, we plot the fidelity against the relaxation time $\tau$ for
two level system with ground state non-degenerate and excited state two-fold
degenerate.
It is easy to see
that the fidelity depends on the ratio of coupling density and energy gap
sensitively. In fact, the perturbation theory requires that the density matrix elements
are much smaller than the energy gap. If the ratio is $10^{-3}$ or $10^{-4}$,
the fidelity is quite close to 1 and shift of final state from
the target state is minor. For ratio is about $10^{-3}$, we
find that the fidelity is getting smaller to 95\% when the relaxation
time is $10^{-12}$s.

\section{Control without system excitation}\label{section-control-2}

In previous section, we presented a scheme of control of degenerate system
through eliminating degeneracy by constant excitation field. We naturally ask whether
we can control the degenerate systems directly without removing the degeneracy.
In this section, we shall address this issue.

Define the energy gap $\mu_i = E_{i+1}-E_i>0$ ($i=1, 2, \cdots,N-1$)
hereafter.


We start by proving the following lemma.

\begin{lemma} \label{lemma1}
If the condition
\be
d_{n1,n+11}d_{n2,n+12}\neq d_{n1,n+12}d_{n2,n+11}, \quad (2\leq n\leq N-1),
\label{condition1}
\ee
is satisfied, and $x_{11,21}, x_{11,22} \in L_0$ or
$y_{11,21}, y_{11,22} \in L_0$,
then $L_0=\su (2N-1)$ when $N\geq 3$ and the system is completely controllable.
\end{lemma}

\noindent {\bf Proof}. From $x_{11,21}, x_{11,22} \in L_0$, we obtain
\beqn
&& \mu_1^{-1}\left[ iH_0, x_{11,21}\right]=y_{11,21}\in L_0, \nonumber \\
&& \mu_1^{-1}\left[ iH_0, x_{11,22}\right]=y_{11,22}\in L_0, \nonumber \\
&& -2^{-1}\left[ x_{11,21}, y_{11,21}\right]= h_{11,21}\in L_0, \nonumber \\
&& -2^{-1}\left[ x_{11,22}, y_{11,22}\right]= h_{11,22}\in L_0,
\eeqn
and
\beqn
&& V_0=iH_0-(d_{11,21}x_{11,21} +d_{11,22}x _{11,22}) \nonumber \\
&& \ \ \  =  \sum_{n=2}^{N-1}
\sum_{p=1}^2 \sum_{k=1}^2 d_{np,n+1k} x_{np,n+1k} \in L_0. \label{28}
\eeqn
Then we have
\beqn
&& [h_{11,21},V_0]=d_{21,31}y_{21,31}+d_{21,32}y_{21,32}\in L_0,   \nonumber\\
&& [h_{11,22},V_0]=d_{22,31}y_{22,31}+d_{22,32}y_{22,32}\in L_0,   \nonumber\\
&& -\mu_2^{-1} \left[iH_0, d_{21,31} y_{21,31}+d_{21,32} y_{21,32}\right]\nonumber\\
&& \ \ \ \ \ \ \ \ = d_{21,31}x_{21,31}+d_{21,32}x_{21,32}\in L_0, \label{lemma11}\\
&& -\mu_2^{-1} \left[iH_0, d_{22,31} y_{22,31}+d_{22,32}y_{22,32}\right]\nonumber \\
&& \ \ \ \ \ \ \ \ = d_{22,31}x_{22,31}+d_{22,32}x_{22,32}\in L_0.\label{lemma12}
\eeqn
On the other hand, we also have
\beqn
&& -\left[x_{11,21}, x_{11,22}\right]=y_{21,22}, \nonumber \\
&& -\left[ y_{21,22}, d_{21,31}x_{21,31}+d_{21,32}x_{21,32}\right]
   = d_{21,31}x_{22,31}+d_{21,32}x_{22,32}, \label{lemma13}\\
&& \left[ y_{21,22}, d_{22,31}x_{22,31}+d_{22,32}x_{22,32}\right]
   = d_{22,31}x_{21,31}+d_{22,32}x_{21,32}.  \label{lemma14}
\eeqn
From Eqs.(\ref{lemma11},\,\ref{lemma14}), under the condition
$d_{21,32}d_{22,31}\neq d_{21,31}d_{22,32}$, we find that
$x_{21,31}, x_{21,32} \in L_0$,
and similarly, from Eqs.(\ref{lemma12}, \ref{lemma13}),
we have $x_{22,31}, x_{22,32} \in L_0$.
Repeating this process for $N-2$ times, we can finally have
$x_{np,n+1k} \in L_0$ $(1\leq n\leq N-1; p,k=1,2)$, and
$y_{np,n+1k} \in L_0$ $(1\leq n\leq N-1; p,k=1,2)$ by evaluating
their commutation relation with $iH_0$. Those operators
generate the Lie algebra $L_0=\su(2N-1)$.

\subsection{The controllability of two-level system}

Let us first consider the simplest case, the two-level system
with non-degenerate ground state and two-fold degenerate excited
state. The interaction Hamiltonian of the system and control field
can be written as
\be
iH_I=d_{1}x_{11,21}+d_{2}x_{11,22},
\ee
where we have written $d_1\equiv d_{11,21}$ and $d_2 \equiv d_{11,22}$ for convenience.
The dynamical
Lie algebra $L_0$ is generated by $iH_0$ and $iH_1$.
It is easy to see that
\begin{eqnarray}
& \left[iH_{0},iH_{I}\right]=\mu iH_{I}^{1}, \\
& iH_{I}^{1}=d_{1}y_{11,21} + d_{2} y_{11,22} \in L_0,
\end{eqnarray}
where $\mu =E_{2}-E_{1}$. Commutation relations
\begin{eqnarray}
&\left[iH_{I},iH_{I}^{1}\right]=2iH_{I}^{2}\in L_0, \label{10} \\
&iH_{I}^{2}= d_{1}^{2}h_{21,11}+
  d_{1}d_{2} x_{21,22}+
  d_{2}^{2}h_{22,11},
\end{eqnarray}
give rise to a new Lie algebra element $iH_I^2$. Four elements
$\left\{iH_0, iH_I, iH_I^1, iH_I^2\right\}$ span an $4$-dimensional
Lie algebra with the following closed Lie product
\begin{eqnarray}
&\left[iH_{0}, iH_{I}\right]=\mu iH_{I}^1, \ \ \
 \left[iH_{0}, iH_{I}^{1}\right]=-\mu iH_{I}, \ \ \
 \left[iH_{0}, iH_{I}^{2}\right]=0, \\
&\left[iH_{I}, iH_{I}^{2}\right]=-2(d_1^2+d_2^2)iH_I^1, \ \ \
\left[iH_{I}^1, iH_{I}^{2}\right]= 2(d_1^2+d_2^2) iH_I.\label{13}
\end{eqnarray}
From (\ref{10}) and (\ref{13}), we find that $\left\{ iH_I, iH_I^1, iH_I^2\right\}$
span an $3$-dimensional Lie algebra. Define
\begin{equation}\fl\qquad
ix=\frac{1}{\sqrt{2(d_1^2+d_2^2)}}iH_I, \ \ \
iy=\frac{1}{\sqrt{2(d_1^2+d_2^2)}}iH_I^1, \ \ \
iz=\frac{1}{\sqrt{2(d_1^2+d_2^2)}}iH_I^2,
\end{equation}
we find the standard commutation relations of $\su(2)$
\begin{equation}
[ix, iy]=iz, \ \ \
[iy, iz]=ix, \ \ \
[iz, ix]=iy.
\end{equation}
Defining a new element $ih_0$ to replace $iH_0$
\begin{equation}
ih_0 \equiv iH_0 -\frac{\mu}{2(d_1^2+d_2^2)} iH_I^2,
\end{equation}
we can check that
\begin{equation}
[ih_0, ix]=[ih_0, iy]=[ih_0, iz]=0.
\end{equation}
Therefore, for 2-level degenerate system, the Lie algebra
generated by $iH_0$ and $iH_I$ is $\su(2)\oplus \mbox{u}(1)$, not
$\su(3)$, namely, the system is not completely controllable.

\subsection{Three level system}

For explicitness, we consider the three-level system in this
subsection and we will see that this system is completely
controllable if some conditions are satisfied. The
system Hamiltonian and the interaction Hamiltonian can
be explicitly written as
\begin{eqnarray}
H_0 = & E_1e_{11,11}+E_2e_{21,21}+E_2e_{22,22}+E_3e_{31,31}+E_3e_{32,32},\nonumber \\
iH_1= & d_{11,21}\, x_{11,21}+d_{11,22}\, x_{11,22}+d_{21,31}\, x_{21,31}+d_{21,32}\, x_{21,32}\nonumber\\
&+d_{22,31}\, x_{22,31}+ d_{22,32}\, x_{22,32}.
\end{eqnarray}
It is easy to find that
\begin{eqnarray}
\tilde{V}_1&=&-\left[iH_0,\left[iH_0,iH_1\right]\right]-\mu_{2}^2 (iH_1) \nonumber\\
&=&\left(\mu_1^2-\mu_2^2\right)\left(d_{11,21}x_{11,21}+d_{11,22}x_{11,22}\right)\in L_0.
\label{daxia}
\end{eqnarray}
Let us consider two different cases.

\vspace{0.2cm}

\noindent {\it Case I:} $\mu_1^2\neq \mu_2^2$. In this case, from
(\ref{daxia}), we can obtain that
\begin{eqnarray}
& V_1^{\prime}=d_{11,21}\, x_{11,21}+d_{11,22}\, x_{11,22} \in L_0, \label{47}\\
& V_1=\mu_1^{-1}[iH_0,V_1^{\prime}]=d_{11,21} \, y_{11,21}+d_{11,22} y_{11,22} \in L_0. \label{46}
\end{eqnarray}
It should be noted that $V_1,V_1^{\prime}$ and
\beqn
V_0=& \frac{1}{2}[V_1,V_1^{\prime}]=i(d_{11,21}^2+d_{11,22}^2)e_{11,11}-
id_{11,21}^2e_{21,21} \nonumber \\
& -id_{11,22}^2e_{22,22}-d_{11,21}d_{11,22}\,  x_{21,22},
\eeqn
generate the Lie
algebra $\su(2)$. For the two-level system discussed in previous subsection,
except $iH_0$, we cannot generate any other elements as we
cannot separate the sum (\ref{47},\ref{46}) into $x_{11,21}\in L_0$
and $x_{11,22}\in L_0$. But fortunately,
for $N$-level system ($N\geq 3$), this can be achieved with the help of terms $x_{2i,3j}$ in $iH_1$.
To see this, we can evaluate the commutator of $V_1$ with $iH_1-V_1^{\prime}$
\beqn
V_2&=&[V_1,iH_1-V_1^{\prime}]=
(d_{11,21}d_{21,31}+d_{11,22}d_{22,31}) \, y_{11,31} \nonumber \\
& & + ((d_{11,21}d_{21,32}+d_{11,22}d_{22,32}) \, y_{11,32},
\eeqn
and
\beqn
V_3&= &
\mu_1^{-1}[iH_0,[V_2,iH_1-V_1^{\prime}]]\nonumber \\
&= &
(pd_{21,31}+qd_{21,32})\, y_{11,21}  +(pd_{22,31}+qd_{22,32}) \, y_{11,22}, \label{49}
\eeqn
where
\be
p=d_{11,21}d_{21,31}+d_{11,22}d_{22,31},\quad q=d_{11,21}d_{21,32}+d_{11,22}d_{22,32}.
\ee
From (\ref{46}) and (\ref{49}), we obtain $y_{11,21}, y_{11,22}\in L_0$,
when
\be
d_{11,21}(pd_{22,31}+qd_{22,32})\neq d_{11,22}(pd_{21,31}+qd_{21,32}).
\ee
If we further require $d_{21,31}d_{22,32}\neq d_{21,32}d_{22,31}$,
from Lemma \ref{lemma1}, we conclude that the 3-level system with
different energy gaps is
completely controllable.

\vspace{2mm}
\noindent {\it Case II: $\mu_1=\mu_2=\mu$}.
\vspace{2mm}

\noindent In this case, Eq.\,(\ref{daxia}) is vanishing and no new element
is generated. We can verify that
\beqn
\fl\qquad
V_1\equiv \mu^{-1}\left[iH_0, iH_1\right]=d_{11,21} y_{11,21}+d_{11,22} y_{11,22}+d_{21,31} y_{21,31}  \nonumber \\
    +d_{21,32}y_{21,32} +d_{22,31}y_{22,31} +d_{22,32}y_{22,32}\in L_0 ,  \\
\fl\qquad V_0 =
\frac{1}{2}[V_1,iH_1]=i(d_{11,21}^2+d_{11,22}^2)e_{11,11}
 +i(d_{21,31}^2+d_{21,32}^2-d_{11,21}^2)e_{21,21}\nonumber \\
 +i(d_{22,31}^2  +d_{22,32}^2-d_{11,21}^2)e_{22,22}
 -i(d_{21,31}^2+d_{22,31}^2)e_{31,31}\nonumber \\
 -i(d_{21,32}^2+d_{22,32}^2)e_{32,32}\nonumber \\
 -(d_{11,21}d_{11,22}-d_{21,31}d_{22,31}-d_{21,32}d_{22,32})x_{21,22} \nonumber \\
 -(d_{21,31}d_{21,32}+d_{22,31}d_{22,32})x_{31,32} \in L_0.
\eeqn
From $V_0$ and $V_1$, we have
\beqn
\fl\qquad
V_2 = \mu^{-1}[iH_0,[V_1,V_0]]=d_{11,21}^{(2)} \, y_{11,21} +d_{11,22}^{(2)} \, y_{11,22} \nonumber \\
\fl\qquad\qquad + d_{21,31}^{(2)} y_{21,31}+
d_{21,32}^{(2)} \, y_{21,32}+
d_{22,31}^{(2)} \, y_{22,31}+
d_{22,32}^{(2)} \, y_{22,32}\in L_0,
\eeqn
where the coefficients $d^{(2)}$ satisfy the following equations
\beqn
\fl\qquad \left(
  \begin{array}{c}
  d_{11,21}^{(2)}\\
  d_{11,22}^{(2)}
  \end{array}\right)
=G_1
\left(
  \begin{array}{c}
  d_{11,21}\\
  d_{11,22}
  \end{array}\right),
  \quad
  G_1 =\left(
  \begin{array}{cc}
  \upsilon_{11,21} & -b_1\\
  -b_1 & \upsilon_{11,22}
  \end{array}\right), \label{57} \\
\fl\qquad
\left(
  \begin{array}{c}
  d_{21,31}^{(2)}\\
  d_{21,32}^{(2)}\\
  d_{22,31}^{(2)}\\
  d_{22,32}^{(2)}
  \end{array}\right)
= G_2
\left(
  \begin{array}{c}
  d_{21,31}\\
  d_{21,32}\\
  d_{22,31}\\
  d_{22,32}
  \end{array}\right),
  \ \
  G_2 = \left(
  \begin{array}{cccc}
  \upsilon_{21,31} & -b_2 & b_1 & 0\\
  -b_2 & \upsilon_{21,32} & 0 & b_1\\
  b_1 & 0 & \upsilon_{22,31} & -b_2\\
  0 & b_1 & -b_2 & \upsilon_{22,32}
  \end{array}\right).
\eeqn
and the parameters $\upsilon_{ij,kl}$ and $b_i$ are defined as
\beqn
& \upsilon_{11,21}=d_{21,31}^2+d_{21,32}^2-2d_{11,21}^2-d_{11,22}^2,\nonumber \\
& \upsilon_{11,22}=d_{22,31}^2+d_{22,32}^2-d_{11,21}^2-2d_{11,22}^2, \nonumber \\
& \upsilon_{21,31}=d_{11,21}^2-2d_{21,31}^2-d_{21,32}^2-d_{22,31}^2,\nonumber \\
& \upsilon_{21,32}=d_{11,21}^2-d_{21,31}^2-2d_{21,32}^2-d_{22,32}^2,\nonumber \\
& \upsilon_{22,31}=d_{11,22}^2-d_{21,31}^2-2d_{22,31}^2-d_{22,32}^2,\nonumber \\
& \upsilon_{22,32}=d_{11,22}^2-d_{21,32}^2-d_{22,31}^2-2d_{22,32}^2, \nonumber \\
& b_1=d_{11,21}d_{11,22}-d_{21,31}d_{22,31}-d_{21,32}d_{22,32}, \nonumber \\
& b_2=d_{21,31}d_{21,32}+d_{22,31}d_{22,32}.
\eeqn
As $G_1$ and $G_2$ are real symmetric matrices, we can diagonalize them by unitary
transformation $U_1, U_2$, respectively
\begin{eqnarray}
U_1 G_1 U_1^{-1} =
&&\left(
  \begin{array}{cc}
  \lambda_{11} & 0\\
  0 & \lambda_{12}
  \end{array}\right), \label{60}\\
U_2 G_2 U_2^{-1} =
&&\left(
  \begin{array}{cccc}
  \lambda_{21} & 0 & 0 & 0\\
  0 & \lambda_{22} & 0 & 0\\
  0 & 0 & \lambda_{23} & 0\\
  0 & 0 & 0 & \lambda_{24}
  \end{array}\right),
\end{eqnarray}
where $\lambda_{1k},\lambda_{2p}$ are eigenvalues of $G_1$ and $G_2$, respectively.
Introducing a set of new parameters
\beqn
\left(
  \begin{array}{c}
  C_{11,21}^{(2)}\\
  C_{11,22}^{(2)}
  \end{array}\right)
=U_1\left(
     \begin{array}{c}
     d_{11,21}^{(2)}\\
     d_{11,22}^{(2)}
     \end{array}\right), \ \ \
\left(
   \begin{array}{c}
   C_{11,21}\\
   C_{11,22}
   \end{array}\right)
=U_1\left(
    \begin{array}{c}
    d_{11,21}\\
    d_{11,22}
    \end{array}\right),
 \\
\left(
   \begin{array}{c}
   C_{21,31}^{(2)}\\
   C_{21,32}^{(2)}\\
   C_{22,31}^{(2)}\\
   C_{22,32}^{(2)}
   \end{array}\right)
=U_2\left(
      \begin{array}{c}
      d_{21,31}^{(2)}\\
      d_{21,32}^{(2)}\\
      d_{22,31}^{(2)}\\
      d_{22,32}^{(2)}
      \end{array}\right), \ \
\left(
  \begin{array}{c}
  C_{21,31}\\
  C_{21,32}\\
  C_{22,31}\\
  C_{22,32}
  \end{array}\right)
=U_2\left(
     \begin{array}{c}
     d_{21,31}\\
     d_{21,32}\\
     d_{22,31}\\
     d_{22,32}
     \end{array}\right),
\eeqn
we can easily obtain that
\beqn
C_{11,21}^{(2)}=\lambda_{11}C_{11,21}, \quad
C_{11,22}^{(2)}=\lambda_{12}C_{11,22}, \quad
C_{21,31}^{(2)}=\lambda_{21}C_{21,31}, \nonumber \\
C_{21,32}^{(2)}=\lambda_{22}C_{21,32}, \quad
C_{22,31}^{(2)}=\lambda_{23}C_{22,31}, \quad
C_{22,32}^{(2)}=\lambda_{24}C_{22,32}.
\eeqn
In terms of those new parameters, $V_1$ can be rewritten as
\beqn
\fl \qquad
V_1= \left(y_{11,21}, y_{11,22}\right)
       {d_{11,21}\choose d_{11,22}}
       +\left(y_{21,31},y_{21,32},y_{22,31},y_{22,32}\right)
        \left(\begin{array}{c}
        d_{21,31} \\ d_{21,32}\\ d_{22,31}\\ d_{22,32}
        \end{array}\right) \nonumber \\
\fl\qquad\quad =\left(y_{11,21}, y_{11,22}\right)U_1^{-1}
       {C_{11,21}\choose C_{11,22}}
       +\left(y_{21,31},y_{21,32},y_{22,31},y_{22,32}\right)U_2^{-1}
        \left(\begin{array}{c}
        C_{21,31} \\ C_{21,32}\\ C_{22,31}\\ C_{22,32}
        \end{array}\right) \nonumber \\
\fl\qquad\quad = \left(\tilde{y}_{11,21}, \tilde{y}_{11,22}\right)
       {C_{11,21}\choose C_{11,22}}
       +\left(\tilde{y}_{21,31},\tilde{y}_{21,32},\tilde{y}_{22,31},\tilde{y}_{22,32}\right)
        \left(\begin{array}{c}
        C_{21,31} \\ C_{21,32}\\ C_{22,31}\\ C_{22,32}
        \end{array}\right) \nonumber \\
\fl\qquad\quad =C_{11,21}\tilde{y}_{11,21}+C_{11,22}\tilde{y}_{11,22}+C_{21,31}\tilde{y}_{21,31}+C_{21,32}\tilde{y}_{21,32}\nonumber\\
\fl\qquad\qquad +C_{22,31}\tilde{y}_{22,31}+C_{22,32}\tilde{y}_{22,32},
\label{61}
\eeqn
where
\beqn
&&\left(\tilde{y}_{11,21}, \tilde{y}_{11,22}\right)=
\left(y_{11,21}, y_{11,22}\right)U_1^{-1},  \\
&&\left(\tilde{y}_{21,31},\tilde{y}_{21,31},\tilde{y}_{21,31},\tilde{y}_{21,31}\right)
=\left(y_{21,31},y_{21,31},y_{21,31},y_{21,31}\right)U_2^{-1}.
\eeqn
Similarly, we have
\beqn
V_2&=&C_{11,21}^{(2)}\tilde{y}_{11,21}+C_{11,22}^{(2)}\tilde{y}_{11,22}+C_{21,31}^{(2)}\tilde{y}_{21,31}\nonumber\\
&&+C_{21,32}^{(2)}\tilde{y}_{21,32}+C_{22,31}^{(2)}\tilde{y}_{22,31}+C_{22,32}^{(2)}\tilde{y}_{22,32}
\nonumber \\
&=&\lambda_{11}C_{11,21}\tilde{y}_{11,21}+\lambda_{12}C_{11,22}\tilde{y}_{11,22}+\lambda_{21}C_{21,31}\tilde{y}_{21,31}\nonumber\\
&&+\lambda_{22}C_{21,32}\tilde{y}_{21,32}+\lambda_{23}C_{22,31}\tilde{y}_{22,31}
+\lambda_{24}C_{22,32}\tilde{y}_{22,32},\label{62}
\eeqn
Observing the Eqs.(\ref{61},\ref{62}), both elements are linear combition of
six elements $\tilde{y}_{ij,kl}$. In the following we always require that
\begin{equation}
\lambda_{11} \neq 0,\quad \lambda_{12}\neq 0,\quad\lambda_{11}\neq \lambda_{12},\quad
C_{11,21}\neq 0,\quad C_{11,22}\neq 0.\label{63}
\end{equation}
If all other $\lambda_{2k} \ (k=1,2,3,4)$ are vanishing, we can find that
\begin{equation}
\tilde{y}_{11,21}\in L_0,\quad \tilde{y}_{11,22}\in L_0. \label{lie2}
\end{equation}
If some $\lambda_{2k}\neq 0$, we need more than Lie elements, which can be
obtained by evaluating the following commutation relations
$V_m = \mu^{-1}\left[iH_0,\left[V_{m-1},V_0\right]\right]$. It is easy to
check that
\beqn
\fl\qquad V_m
=\lambda_{11}^{m-1}C_{11,21}\tilde{y}_{11,21}+\lambda_{12}^{m-1} C_{11,22} \tilde{y}_{11,22}
+\lambda_{21}^{m-1}C_{21,31}\tilde{y}_{21,31}\nonumber\\
\fl\qquad\qquad
+\lambda_{22}^{m-1}C_{21,32}\tilde{y}_{21,32}+\lambda_{23}^{m-1}C_{22,31}\tilde{y}_{22,31}
+\lambda_{24}^{m-1}C_{22,32}\tilde{y}_{22,32}\in L_0.
\eeqn
We then obtain $M$ those type of elements by choosing $m=1,\cdots,M$, where
$M$ is the number of non-zero $\lambda_{ij,kl}$'s. The coefficient matrix of those
element is Vandermonde's matrix with non-zero determinant when all non-zero
$\lambda$'s are different from any others. If this condition is satisfied,
we again obtain the Lie algebra element (\ref{lie2}).
As the transformation matrix $U_1$ is nonsingular,
we obtain $y_{11,21}, y_{11,22} \in L_0$.
In conclusion, the system with equal energy gaps is completely controllable
if conditions (\ref{63}) (\ref{condition1}) are satisfied and all non-zero $\lambda_{ij}$'s
are distinctive.

We remark that if all $\lambda_{ij} \neq 0$ and
$C_{ij,kl} \neq 0$, we can obtain all six elements $y_{ij,kl}\in L_0$
from six elements $V_m$ and then the generated Lie algebra is SU(5). In
this case, we do not need to use Lemma 1 and thus
the condition (\ref{condition1}) is unnecessary.

\subsection{Controllability of systems with different energy gaps}

Now let us turn to the control of arbitrary $N$-dimensional quantum
systems. We first consider in this subsection the system with
distinct energy gaps.

\begin{theorem}
If $\mu_1 \neq \mu_n, n=2,3,...,N-1$, and the coupling
constants satisfy the following conditions
\be
d_{11,21}(pd_{22,31}+qd_{22,32})\neq d_{11,22}(pd_{21,31}+qd_{21,32}), \label{24}
\ee
where $p=d_{11,21}d_{21,31}+d_{11,22}d_{22,31}$,
$q=d_{11,21}d_{21,32}+d_{11,22}d_{22,32}$,
we have $y_{11,21}$, $y_{11,22} \in L_0$ and thus the system is completely
controllable.
\end{theorem}

\noindent{\bf Proof}. We evaluate
\beqn
&&V'=[iH_0,iH_1]=\sum_{n=1}^{N-1} \sum_{j=1}^{\beta_n} \sum_{k=1}^2 \mu_n d_{nj,n+1k}\, y_{nj,n+1k}\in L_0,   \nonumber\\
&&V''=-[iH_0,V']=\sum_{n=1}^{N-1} \sum_{j=1}^{\beta_n} \sum_{k=1}^2 \mu_n^2d_{nj,n+1k}\, x_{nj,n+1k}\in L_0.
\eeqn
From $V^\prime$ and $V^{\prime\prime}$, we have
\beqn
V_1
& = &  V^{\prime\prime}-\mu_{N-1}^2 (iH_1)\nonumber \\
& = &  \sum_{n=1}^{N-2} \sum_{j=1}^{\beta_n}
\sum_{k=1}^2 (\mu_n^2-\mu_{N-1}^2)d_{nj,n+1k} x_{nj,n+1k} \in L_0.
\eeqn
Repeating this process for $iH_0$ and $V_1$ leads to
\beqn
V_2&=&-[iH_0,[iH_0,V_1]]-\mu_{N-2}^2V_1,   \nonumber\\
&=&\sum_{n=1}^{N-3} \sum_{j=1}^{\beta_n} \sum_{k=1}^2 (\mu_n^2-\mu_{N-2}^2)(\mu_n^2-\mu_{N-1}^2)d_{nj,n+1k}x_{nj,n+1k} \in L_0.
\eeqn
After $N-2$ iterations, we have
\be
V_{N-2}=\left[\prod_{n=1}^{N-1} (\mu_1^2-\mu_n^2)\right](d_{11,21}x_{11,21}+d_{11,22}
x_{11,22})\in L_0.
\ee
Since $\mu_1$ is different from any others,
$\prod_{n=1}^{N-1} (\mu_1^2-\mu_n^2)\neq0$. This means that
\be
V^\prime_{N-2}=d_{11,21}x_{11,21}+ d_{11,22} x_{11,22}\in L_0, \label{71}
\ee
and
\be
V^{\prime\prime}_{N-2}=\mu_1^{-1}[iH_0,V'_{N-2}]
=d_{11,21}\, y_{11,21}+d_{11,22}\, y_{11,22}\in L_0.
\ee
To prove the complete controllability, we need to separate the
Eq.\,(\ref{71}) to get $x_{11,21}, x_{11,21}\in L_0$. Let us
discuss this in two different cases.

{\it Case 1}: $N=3$. In this case, the commutator
between $V^{\prime\prime}_{N-2}$ and $V_0$ is
\beqn
X&=&[[V^{\prime\prime}_{N-2},V_0],V_0]   \nonumber\\
&=&(pd_{21,31}+qd_{21,32})x_{11,21}+(pd_{22,31}+qd_{22,32})x_{11,22}.
\label{relation2}
\eeqn
where $V_0$ is given in Eq.\,(\ref{28}).
From (\ref{71}) and (\ref{relation2}), we know that, if condition
(\ref{24}) is satisfied, $x_{11,21}, x_{11,21}\in L_0$. Then from
Lemma \ref{lemma1}, the system is completely controllable.

{\it Case 2}: $N>3$. In this case, the commutator
\beqn
X&=&[[V^{\prime\prime}_{N-2},V_0],V_0]   \nonumber\\
&=&(pd_{21,31}+qd_{21,32})x_{11,21}+(pd_{22,31}+qd_{22,32})x_{11,22}+  \nonumber\\
&& (pd_{31,41}+qd_{32,41})x_{11,41} +(pd_{31,42}+qd_{32,42})x_{11,42}\in L_0,
\eeqn
has two additional terms $x_{11,41}$ and $x_{11,42}$.
Fortunately, we can remove those terms by evaluating
\beqn
X^\prime &=&[\mu_1^2-(\mu_1+\mu_2+\mu_3)^2]^{-1} \Big([iH_0,[iH_0,X]]-(\mu_1+\mu_2+\mu_3)^2 X\Big)   \nonumber\\
&=&(pd_{21,31}+qd_{21,32})x_{11,21}+(pd_{22,31}+qd_{22,32})x_{11,22}\in L_0. \label{68}
\eeqn
From Eqs.(\ref{71},\ref{68}), and the condition (\ref{24}),
we obtain $x_{11,21}, x_{11,22}\in L_0$,
and moreover, $y_{11,21}, y_{11,22} \in L_0$.
The conclusion then follows from Lemma 1.

\subsection{Complete controllability of systems with equal energy gaps}

In this subsection we consider the system with equally spaced energy
gaps, namely $\mu\equiv \mu_1=\mu_2=\cdots=\mu_{N-1}\neq 0$.
Let us first introduce some parameters
\beqn
&& K_{ij,ij}^2=\left\{
\begin{array}{ll}
    d_{11,21}^2+d_{11,22}^2, &  i=1, j=1;\\
    \sum\limits_{\alpha=1}^{\beta_{i+1}}
    d_{ij,i+1\alpha}^2-\sum\limits_{\gamma=1}^{\beta_{i-1}}
    d_{i-1\gamma,ij}^2, &  2\leq i\leq N-1; j=1,2; \\
    -(d_{N-11,Nj}^2+d_{N-12,Nj}^2), &  i=N; j=1,2.
\end{array}
\right. \nonumber \\
&& \nu_{ij,i+1k}=K_{i+1K,i+1K}^2-K_{ij,ij}^2,
\eeqn
and
\beqn
b_1 =  d_{11,21}d_{11,22}-d_{21,31}d_{22,31}-d_{21,32}d_{22,32},\nonumber \\
b_i =  d_{i1,i+11}d_{i1,i+12}+d_{i2,i+11}d_{i2,i+12}-d_{i+11,i+21}d_{i+12,i+21}\nonumber \\
\ \ \ \ \ \       -d_{i+11,i+22}d_{i+12,i+22}, \ \ \  2\leq i\leq N-2; \nonumber\\
b_{N-1} =   d_{N-11,N1}d_{N-11,N2}+d_{N-12,N1}d_{N-12,N2}.
\eeqn

Let $V=iH_1$. We first observe that
\be
\widetilde{V}=\mu^{-1}[iH_0,iH_1]=\sum_{i=1}^{N-1} \sum_{j=1}^{\beta_i}
\sum_{k=1}^2 d_{ij,i+1k} y_{ij,i+1k}\in L_0.
\ee
Sum and difference of $\widetilde{V}$ with $V$ give rise to
\begin{eqnarray}
&& V_1^+=\sum_{i=1}^{N-1} \sum_{j=1}^{\beta_i} \sum_{k=1}^2 d_{ij,i+1k}(x_{ij,i+1k}+y_{ij,i+1k}),\\
&& V_1^-=\sum_{i=1}^{N-1} \sum_{j=1}^{\beta_i} \sum_{k=1}^2 d_{ij,i+1k}(x_{ij,i+1k}-y_{ij,i+1k}),
\end{eqnarray}
which, along with their commutator
\be
V_1^0=\frac{1}{4}\left[V_1^+,V_1^-\right]=
i\sum_{i=1}^N \sum_{j=1}^{\beta_i} K_{ij,ij}^2 \, e_{ij,ij}-\sum_{i=1}^{N-1} b_i x_{i+1\,1,i+2\,2},
\ee
are all in $L_0$.
Starting from $\widetilde{V}$ and $V_1^0$, we have
\be
\widetilde{V}_1
=\mu^{-1}\left[iH_0,[\widetilde{V},V_1^0]\right]
=\sum_{i=1}^{N-1} \sum_{j=1}^{\beta_i}
\sum_{k=1}^2  d_{ij,i+1k}^{(2)} \, y_{ij,i+1k},
\ee
where we have assumed $b_0=0$ and introduced the
notation
\be
d_{ij,i+1k}^{(2)}\equiv
\nu_{ij,i+1k}d_{ij,i+1\,k}-b_id_{ij,i+1p}+b_{i-1}d_{i\alpha,i+1k},
p\neq k, \alpha\neq j.
\ee
Repeating the process for $n$ times, we obtain the element $\widetilde{V}_n$
\be
\widetilde{V}_n=\sum_{i=1}^{N-1} \sum_{j=1}^{\beta_i}
\sum_{k=1}^2 d_{ij,i+1k}^{(n+1)}\, y_{ij,i+1k}, \label{82}
\ee
where the coefficients satisfy Eq.(\ref{57}) for $i=1$
and the following recursion relations
\be
\left(
\begin{array}{c}
d_{i1,i+11}^{(n+1)}\\
d_{i1,i+12}^{(n+1)}\\
d_{i2,i+11}^{(n+1)}\\
d_{i2,i+12}^{(n+1)}
\end{array}
\right)
=G_i
\left(
\begin{array}{c}
d_{i1,i+11}^{(n)}\\
d_{i1,i+12}^{(n)}\\
d_{i2,i+11}^{(n)}\\
d_{i2,i+12}^{(n)}
\end{array}
\right),
\ee
where
\begin{equation}
G_i = \left(
\begin{array}{cccc}
\nu_{i1,i+11} & -b_i  &  b_{i-1} & 0 \\
-b_i          &  \nu_{i1,i+12} & 0 & b_{i-1}\\
b_{i-1}  &  0  & \nu_{i2,i+11} & -b_i \\
0 & b_{i-1} & -b_i & \nu_{i2,i+12}
\end{array}
\right)
\end{equation}
is independent of $n$.
Noting that the coefficient matrix $G_1$ given in (\ref{57})
and $G_i$
is a real symmetric matrix and can be diagonalized through unitary
transformations $U_1$ (see Eq.\,(\ref{60})) and $U_i$
\be
U_i G_i U_i^{-1} =
\left(\begin{array}{cccc}
    \lambda_{i1} & 0 & 0 & 0\\
    0 & \lambda_{i2} & 0 & 0\\
    0 & 0 & \lambda_{i3} & 0\\
    0 & 0 & 0 & \lambda_{i4}
    \end{array}\right),
\ee
where the diagonal elements $\lambda_{ij}$ ($j=1,2,3,4$) are the eigenvalue of $G_i$.

Introduce a set of new parameters
\begin{equation}
  \left(\begin{array}{c}
     C_{i1,i+11}^{(n+1)}\\
     C_{i1,i+12}^{(n+1)}\\
     C_{i2,i+11}^{(n+1)}\\
     C_{i2,i+12}^{(n+1)}
  \end{array}\right)
 =G_i\left(\begin{array}{c}
     d_{i1,i+11}^{(n+1)}\\
     d_{i1,i+12}^{(n+1)}\\
     d_{i2,i+11}^{(n+1)}\\
     d_{i2,i+12}^{(n+1)}
     \end{array}\right).
\end{equation}
Then we can easily obtain the recurrence relations between these new parameters
\beqn
&& C_{i1,i+11}^{(n+1)}=\lambda_{i1}C_{i1,i+11}^{(n)},\ \ \
C_{i1,i+12}^{(n+1)}=\lambda_{i2}C_{i1,i+12}^{(n)},\nonumber \\
&& C_{i2,i+11}^{(n+1)}=\lambda_{i3}C_{i2,i+11}^{(n)},\ \ \
C_{i2,i+12}^{(n+1)}=\lambda_{i4}C_{i2,i+12}^{(n)}.
\eeqn
We can rewritten the element (\ref{82}) as
\be
\fl\qquad
\widetilde{V}_n=\sum_{i=1}^{N-1} \sum_{j=1}^{\beta_i} \sum_{k=1}^2
C_{ij,i+1k}^{(n+1)} \tilde{y}_{ij,i+1k}=
\sum_{i=1}^{N-1} \sum_{j=1}^{\beta_i} \sum_{k=1}^2
\lambda_{ip_{jk}}^n C_{ij,i+1k}^{(1)} \tilde{y}_{ij,i+1k}
\in L_0,  \label{50}
\label{tildex}
\ee
where $m=1,2,...,M$,
$p_{11}=1$, $p_{12}=2$, $p_{21}=3$, $p_{22}=4$, and
\begin{equation}
\fl\qquad  \left(
      \tilde{y}_{i1,i+11},
      \tilde{y}_{i1,i+12},
      \tilde{y}_{i2,i+11},
      \tilde{y}_{i2,i+12}
  \right)
=\left(
      y_{i1,i+11},
      y_{i1,i+12},
      y_{i2,i+11},
      y_{i2,i+12}
      \right)U_i^{-1}.
\end{equation}

Let $M$ be the
number of non-zero $C_{ij,i+1k}^{(1)} \tilde{y}_{ij,i+1k}$ in (\ref{50}).
Then we get a set of equations about
$C_{ij,i+1k}^{(1)} \tilde{y}_{ij,i+1k}$ whose coefficient
matrix is the square Vandermonde's matrix. If the determinant of
coefficient matrix is not vanishing, or in other words,
all $\lambda_{ij}$ are different from any others, we can obtain
$
C_{11,21}^{(1)} \tilde{y}_{11,21},
C_{11,22}^{(1)} \tilde{y}_{11,22}\in L_0
$
and further
$
y_{11,21},
y_{11,22}\in L_0
$
due to $
C_{11,21}^{(1)},
C_{11,22}^{(1)}\neq 0
$
and the unitarity of the matrix $U_i$.
From Lemma {\ref{lemma1}}, we conclude that $L_0=\su(2N-1)$
and the system is completely controllable.

In summary, we conclude that

\begin{theorem}
The degenerate quantum system with $N$ equally spaced energy levels
is completely controllable if the parameters satisfy the
following conditions
\begin{itemize}
\item Condition (\ref{condition1}) for the validity of Lemma 1;
\item Condition (\ref{63});
\item All non-zero $\lambda_{ij}$ are different from any others to guarantee
determinant of Vandermonde's matrix non-vanishing.
\end{itemize}
\end{theorem}

As an explicit example, one can check that
the degenerate system with $E_n=n-1/2$ and
$d_{ij,i+1k}=(N+3-i-j-k)^{1/2}$ is completely
controllable.

\section{Conclusion}

In this paper, we have systematically investigated the control of quantum
system with energy degeneracy using two different approach.
The first approach is to apply a weak constant field to eliminate the
degeneracy and then control it using techniques developed for
non-degenerate quantum system. We first examine the conditions
for the elimination of degeneracy and then address the issue
of influence of relaxation time of constant external field to
the target state by calculating the fidelity.

We then investigate the completely controllability of degenerate
system by a single control field only. It is found that the
two level system is not completely controllable in this
control scheme. But fortunately, the multi-level system with
more than two energy levels are completely controllable if
the energy gap and the value of the transition dipole moments
$d_{nk,n+1p}$ satisfy some conditions. Two different cases, namely
the system with different energy gaps and with equal energy
gaps, are considered as in the non-degenerate case.

In the forthcoming papers we shall consider the control the
quantum system with general degeneracy degree rather than
just 2 in this paper. Such investigation might find applications
in the control of molecular system and chain of qubits in quantum
computation.

\section*{Acknowledgement}

This work is supported in part by NFRPC 973 Project under grand number
2006CB921205.


\vspace{1cm}


\begin{thebibliography}{99}
\bibitem{1} 1987 {\it Information Complexity and Control in Quantum Physics}
ed A Blaquiere, S Dinerand and G Lochak (New York: Springer)

\bibitem{2} Butkovskiy A G and Samoilenko Yu I 1990 {\it Control of
Quantum-Mechanical Processes and Systems} (Dordrecht: Kluwer)

\bibitem{3} Jurdjevic V 1997 {\it Geometric Control Theory}
(Cambridge: Cambridge University Press)

\bibitem{4} Lloyd S 2000 {\it Phys. Rev.} A {\bf 62} 022108

\bibitem{huang} Huang G M, Tarn T J and Clark J W 1983
{\it J. Math. Phys.} {\bf 24} 2608

\bibitem{controllability}
Ramakrishna V and Rabitz H 1996 {\it Phys. Rev.} A {\bf 54} 1715

\bibitem{chem} Rabitz H, de Vivie-Riedle R, Motzkus M and Kompa K 2000 {\it Science} {\bf 288} 824

\bibitem{solo} Schirmer S G, Solomon A I and Leahy J V 2002 {\it J. Phys. A} {\bf 35} 4125

\bibitem{fu1} Fu H, Schirmer S G and Solomon A I 2001 {\it J. Phys. A: Math. Gen.} {\bf 34} 1679

\bibitem{fu2} Schirmer S G, Fu H and Solomon A I 2001 {\it Phys. Rev. A} {\bf 63} 063410



\bibitem{graph1} Turinici G 2000 {\it Mathematical Models and Methods for ab Initio Quantum Chemistry} (Lecture Notes in
Chemistry vol 74) ed M Defranceschi and C Le Bris (Berlin: Springer)

\bibitem{graph2} Turinici G and Rabitz H 2001 {\it Chem. Phys.} {\bf 267} 1

\bibitem{graph3} Rangan C and Bloch A M 2005 {\it J. Math. Phys.} {\bf 46} 032106

\bibitem{cab} Cabrera R, Bayis W E, Rangan C 2007 {\it Phys. Rev. A.} {\bf 76} 033401 \\
Albertini F and D'Alessandro D 2003 {\it IEEE Trans. Autom. Control} {\bf 48} 1399

\bibitem{liec} Brockett R W 1973 {\it SIAM J. Appl. Math.} {\bf 25} 213\\
Brockett R W 1973 {\it Lie Algebras and Lie Groups in Control Theory Geometric Methods in System Theory}
ed D Q Mayne and R W Brockett (Dordrecht: Reidel) pp 43¨C82

\bibitem{fu3} Fu H C, Dong H, Liu X F and Sun C P 2007 {\it Phys. Rev.} A {\bf 75} 052317\\
Fu H C, Dong H, Liu X F and Sun C P 2009 {\it J. Phys. A: Math. Theor.} {\bf 42} 045303

\bibitem{lie} Humphreys J E 1972 {\it Introduction to Lie Algebras and Representation Theory} (New York: Springer)

\bibitem{quantumbook} Landau L D and Lifshitz E M 1977 {\it Quantum Mechanics} ({\it Non-relativistic Theory}) (Reed
Educational and Professional Publishing Ltd, Third Edition)
\end{thebibliography}
\end{document}